# Nonlinearity and trapping in excitation transfer: Dimers and Trimers.

Ivan Barvík

Institute of Physics, Charles University,

Ke Karlovu 5, 121 16 Prague, Czech Republic

barvik@ns.karlov.mff.cuni.cz

Bernd Esser
and
Holger Schanz

Institute of Physics, Humboldt University

Invalidenstr.110, 10099 Berlin, Germany

itp@itp02.pysik.hu-berlin.de

## Abstract

We study the interplay between nonlinearity in exciton transport and trapping due to a sink site for the dimer and the trimer with chain configuration by a numerical integration of the discrete nonlinear Schrödinger equation.

Our results for the dimer show, that the formation of a self trapped state due to the nonlinear coupling increases the life time of the exciton substantially. Self trapping can be enhanced by the sink for short times, but for long times it disappears. In the trimer consisting of a subdimer extended by a sink site exists a transition between states localized on the two sites of the subdimer before for larger nonlinear coupling self trapping on one site of the subdimer is observed.

For large trapping rates the fear of death effect leads to an increasing life time of the excitation on both, the dimer and the trimer. The sink site is then effectively decoupled. We explain this effect using an asymptotic theory for strong trapping and demonstrate it by direct numerical computation.

# 1  Introduction

The purpose of this paper is to study the interplay between a coherent transfer regime of an exciton and two processes leading to the loss of the linear character of the exciton transfer, namely trapping of the particle and transport nonlinearity. The system under study is a moving quasiparticle interacting strongly with polarization vibrations. At one of the sites of the configuration trapping with a prescribed rate $\gamma$ can occur.

Trapping of quasiparticles constitutes an important phenomenon in many areas of physics. In photosynthesis, for instance, an exciton in a harvesting antenna transfers its energy to a so called reaction center, where it can be trapped. Electron transfer processes then follow.

Much work has been done on the exciton transfer theories in the last two decades. The main effort has been directed to obtaining equations which describe coupled coherent and incoherent motion of the excitation. The corresponding theories include Pauli Master Equations (PME), General Master Equations (GME), Stochastic Liouville equations (SLE) and Continuous Time Random Walk (CTRW) [1–26]. Only recently [17–28] the problem of a rigorous description, leading to positive occupation probabilities, of the trap (modelled as a so called sink) in the coherent and near coherent regimes of the excitation transfer has been solved using the SLE, GME and CTRW methods.

Using the SLE [17–20] we have shown for more extended systems, that in presence of a sink, the excitation behaves as it would like to avoid the site influenced by the sink. This effect has been observed also in the case of a semiinfinite linear chain [21]. We called it the "fear of death effect". In this case the decay of the whole occupation probability P(t) with time is slower for larger $\gamma$.

In particular, in a linear trimer with a sink introduced at one end and for a very large trapping rate $\gamma$, the rest (dimer) could be taken as decoupled. The exciton moves from the very beginning approximately in the dimer avoiding the place influenced by the sink. Analogously, in the hexagonal model of the photosynthetic unit the reaction centre seems to be almost decoupled from the antenna system for a very large trapping rate $\gamma$. Still there is no doubt that after a long time the whole occupation probability P(t) of the exciton disappears due to the trapping term. In our recent paper [27] we investigated in some simple finite systems (dimer, trimer) the interplay between the trapping and the interaction with a bath described in the framework of the SLE with the formal Haken Strobl parametrization.

In the last few years much attention has been paid to the problem of nonlinear interactions in coupled exciton-vibration systems [30–29]. A possible approach to it is the so called discrete nonlinear Schrödinger equation (DNLSE, sometimes also called discrete self trapping equation) which, though not yet rigorously justified has the advantage of allowing for a simple analytical treatment. In this equation besides the transfer matrix element $V$ a nonlinear coupling constant $\chi$ discribing the site energy lowering due to quasiparticle formation is contained.

One encounters the discrete nonlinear Schrödinger equation in the context of numerous phenomena from various fields of physics.

Although the DNLSE was already mentioned in early papers on the polaron problem (see e.g. [52]) and much physics has been extracted through numerical analysis [29] exact solutions are not known in general. However, it has been found recently, that considerable insight can be gained into essential physics of the system through the exact analytical solutions when the system has only two sites. This case of the nonlinear dimer has been investigated during the last view years from many points of view. Kenkre with his collaborators [30–42] and Szöcs and Banacky [43,44] have tackled the problem with the occupation probability difference as the most important quantity in mind. The exact time evolution for the occupation probability difference was given in terms of Jacobian elliptic functions for arbitrary initial conditions. Setting the occupation of the first site initially to one the transition to a self trapped state was obtained by Kenkre et al. [30] in a dimer with $\chi \geq 4V$. However, one sided oscillations in the occupation difference appear for some initial condition already from $\chi \geq 2V$. This can be related to the appearence of a homoclinic structere on the Bloch sphere, which was shown by Esser and Hennig [45–49]. For increasing $\chi$ the separatrix of this homoclinic structure grows and at $\chi = 4V$ starts to include the poles of the sphere which correspond to initially localized states.

The limiting behaviour for $t \to \infty$ has been treated with and without dissipation which was modelled in the framework of the stochastic Liouville equations with diagonal and nondiagonal Haken-Strobl parameters. Attention has been paid also to the influence of symmetry in the dimer, trimer and n-mer.

Excitation transfer in trapless dimers and trimers can be described by Hamiltonian flows located on the surface of the Bloch sphere for the dimer and its higher dimensional analog for the trimer case. This allows to use the methods of nonlinear dynamics in the analysis of the transfer properties, in particular to prove the appearence of bifurcations in the flow line picture [29,45–49]. Esser with his collaborators pointed out another interesting feature: the possibility of chaotic behaviour due to a perturbation of the homoclinic structure which constitutes a route to chaos. In this case one has to leave the description of the problem in terms of occupation probabilities only and must include the evolution of all Bloch variables, i.e. one must follow precisely the time development of the nondiagonal elements of the corresponding density matrix. In this connection interesting results have also been obtained in the nonlinear dimer, when the adiabatic assumption is dropped, i.e. when vibrational relaxation which leads to the nonlinear transport is not assumed to be infinitely fast [50].

The aim of our contribution is to investigate the interplay between the nonlinearity and the trapping of the excitation (here modeled as a sink). We shall follow the influence of both the nonlinearity and trapping on the transfer properties.

The paper is organized as follows. In section II the formulation of the problem and basic equations are given. Sections III and IV contain equations which describe the dynamics of a dimer and a trimer, respectively, including nonlinear coupling and the influence of a sink placed at one end of the configuration. We present their numerical solutions and derive some analytic expressions for the case of very strong trapping. In section V conclusions are drawn.

## 2 Formulation of the model and basic equations

### 2.1 The DST equation

As the basic equation describing the excitation transfer in our systems we use the discrete nonlinear Schrödinger equation

$$i\frac{d}{dt}c_n(t) = (\varepsilon_n - \chi_n|c_n|^2)c_n + \sum V_{nm}c_m \qquad (1)$$

where $c_n$ is the probability amplitude of the excitation (in what follows denoted as exciton) to occupy the molecule at the n-th site, $\varepsilon_n$ the exciton site energy, $V_{nm}$ ($n \neq m$) the transfer matrix element and $\chi_n$ the nonlinearity parameter describing the lowering of the site energy by exciton occupation. Introducing the density matrix elements

$$\rho_{mn}(t) := c_m(t)c_n^*(t) \qquad (2)$$

we describe the system evolution by the density matrix equation

$$i\frac{\partial}{\partial t}\rho_{nm} = (\epsilon_n - \epsilon_m)\rho_{nm} + \sum_k (V_{nk}\rho_{km} - \rho_{nk}V_{km}). \qquad (3)$$

Here is

$$\epsilon_n := \varepsilon_n - \chi_n \rho_{nn} \qquad (4)$$

an effective site energy.

### 2.2 Influence of the sink

Trapping at a particular site $s$, consistent with the condition $\rho_{nn}(t) > 0$ for any time $t$, can be introduced into eq. (3) by supplementing the r.h.s. with the terms

$$-i(\gamma/2)\rho_{ns} \qquad (n \neq s) \qquad (5)$$

for the non-diagonal and

$$-i\gamma\rho_{ss} \qquad (6)$$

for the diagonal matrix elements connected with the trap site $s$.

The description of the influence of the trap as given by the terms (5), (6) is equivalent to introducing into the Hamiltonian the matrix element $H_{ss} = -i\gamma/2$ for the sink place $s$.

## 2.3 Perturbation theory for a large $\gamma$

In the case of a large trapping rate $\gamma$ at site $s$, i.e. when $\gamma$ exceeds by far all the other energies entering the r.h.s of the eq. (3), one can simplify the system (3-6) by using quasistationarity in the equations for matrix elements $\rho_{mn}$ involving the trapping site. This corresponds to neglecting times smaller than $\gamma^{-1}$ in the evolution of the system and is formally achieved by setting $\dot{\rho}_{mn} = 0$ for $m = s$ or $n = s$ ($m \neq n$) and $m = n = s$. This procedure is quite analogous to that used for the transition from the stochastic Liouville to the Pauli Master equation in [3,4] when quasistationarity is assumed for the non-diagonal density matrix elements due to the fast phase relaxation connected with the stochastic sources. Here we apply this approximation to a part of the nondiagonal density matrix elements, namely those connected with the trap site. As we will show, this results in analytical expressions for the fear of death effect in the case of large $\gamma$. One finds for the nondiagonal elements $n \neq s$

$$\rho_{ns} = \frac{1}{\epsilon_n - \epsilon_s - i(\gamma/2)} \sum_k (\rho_{nk} V_{ks} - V_{nk} \rho_{ks}) \tag{7}$$

and for the diagonal elements connected with the trap

$$\rho_{ss} = \frac{i}{\gamma} \sum_k (\rho_{sk} V_{ks} - V_{sk} \rho_{ks}). \tag{8}$$

From the r.h.s. of eq. (7) it is evident that the $\rho_{ns}$ are of the order of the small parameter $(V/\gamma) \ll 1$ where $V$ is the typical value for the transfer matrix elements entering the r.h.s. of (7). Evaluating (7) and (8) in this small parameter iteratively one obtains approximations to $\rho_{ns}$ and $\rho_{ss}$ for the case of large $\gamma$ which complement our numerical results and are considerd below. We shall now treat the dimer and the trimer separately.

## 3 The dimer with a sink

### 3.1 Equations of motion

Restricting the transfer to the two sites of a dimer with $V_{12} = V_{21} = -V < 0$ one arrives at

$$\begin{aligned} i\dot{\rho}_{11} &= +V(\rho_{12} - \rho_{21}) \\ i\dot{\rho}_{22} &= -V(\rho_{12} - \rho_{21}) - i\gamma\rho_{22} \\ i\dot{\rho}_{12} &= +V(\rho_{11} - \rho_{22}) - i(\gamma/2)\rho_{12} + (\epsilon_1 - \epsilon_2)\rho_{12} \\ i\dot{\rho}_{21} &= -V(\rho_{11} - \rho_{22}) - i(\gamma/2)\rho_{21} - (\epsilon_1 - \epsilon_2)\rho_{21} \end{aligned} \tag{9}$$

Passing to the Bloch variables

$$\begin{aligned} x_1 &:= \rho_{12} + \rho_{21} \\ x_2 &:= i(\rho_{12} - \rho_{21}) \\ x_3 &:= \rho_{11} - \rho_{22} \end{aligned} \qquad (10)$$

and introducing the norm as a separate variable

$$n := \rho_{11} + \rho_{22} \qquad (11)$$

one obtains the equations of motion

$$\begin{aligned} \dot{x}_1 &= -(\gamma/2)\,x_1 + \Delta\epsilon(t) x_2 \\ \dot{x}_2 &= -(\gamma/2)\,x_2 - \Delta\epsilon(t) x_1 + 2V x_3 \\ \dot{x}_3 &= -(\gamma/2)\,(x_3 - n) - 2V x_2 \\ \dot{n} &= -(\gamma/2)\,(n - x_3) \end{aligned} \qquad (12)$$

with $\Delta\epsilon(t) := (\varepsilon_2 - \varepsilon_1) + \chi x_3$.

## 3.2 Previously obtained results

In the absence of trapping, i.e. for $\gamma = 0$, the system (12) reduces to the nonlinear Bloch equations considered in [45]. In this case besides the energy there is a second integral of the motion restricting the Bloch variable to the surface of a sphere.

In [45] the fixed points of the trapless system (12) were analyzed. One finds that for $\chi = 2V$ a bifurcation is realized which follows for the symmetric dimer ($\varepsilon_1 = \varepsilon_2$) from $\dot{x}_2 = 0$, i.e.

$$(2V - \chi x_1)x_3 = 0 \qquad (13)$$

For $\chi \leq 2V$ the only solution of eq. (13) is given by $x_3 = 0$, whereas for $\chi > 2V$ the new solution $x_1 = \frac{2V}{\chi}$ appears, resulting in two fixed points with $x_2 = 0$, $x_3 = \pm\sqrt{1 - (\frac{2V}{\chi})^2}$. These points correspond to the new ground states arising from the polaronic effect which holds the quasiparticle preferentially at one of the two sites of the dimer. The solution $x_3 = 0$ for $\chi > 2V$ corresponds to an unstable hyperbolic point at the center of a separatrix. Into this homoclinic structure the new ground states are embeded [45–47].

Kenkre et al. were able to rewrite the system (12) into a closed equation for the probability difference $x_3 = p(t) = P_1(t) - P_2(t)$. The analytical and numerical solution leads for the case of the localized initial condition $P_1(0) = 1$ to the so called self trapped states for $\chi \geq 4V$ [30].

We now turn to investigating the influence of the nonlinearity on the trapping process of the exciton in the dimer. We shall check carefully changes in the dynamic properties due to the trapping rate $\gamma$.

## 3.3 Analytical results for strong trapping

From (7), (8) we know that strong trapping leads to the expressions

$$\rho_{12} = \frac{V}{\epsilon_1 - \epsilon_2 - i(\gamma/2)}(\rho_{22} - \rho_{11}) \qquad (14)$$

$(\rho_{21} = \rho_{12}^*)$ and

$$\rho_{22} = \frac{iV}{\gamma}(\rho_{12} - \rho_{21}) \qquad (15)$$

from which an explicit expression for the occupation of the sink site $\rho_{22}$ can be derived. In the case without nonlinearity ($\chi = 0$) one finds

$$\rho_{22}^0 = \frac{V^2}{(\varepsilon_1 - \varepsilon_2)^2 + (\gamma/2)^2}\rho_{11}^0 \qquad (16)$$

which demonstrates the fear of death effect: due to the strong trapping the occupation of the sink site is very small, namely $\rho_{22} \sim (V/\gamma)^2 \rho_{11} \ll \rho_{11}$. The effect of nonlinearity in the perturbation series for strong trapping is to modify the energy difference in the denominator:

$$\rho_{22} = \frac{V^2}{(\varepsilon_1 - \varepsilon_2 - \chi\rho_{11})^2 + (\gamma/2)^2}\rho_{11} \qquad (17)$$

## 3.4 Numerical results

We have calculated the time dependence of the Bloch variables in presence of nonlinearity and trapping by a direct numerical integration of the nonlinear equation (1) including the trapping as described in Sec. 2.2.

Our results are displayed in figs. 1 - 7 for different trapping rates $\gamma$ and various nonlinearity parameters $\chi$. We have chosen for all figures a symmetric configuration ($\varepsilon_1 = \varepsilon_2$) with $V = 1$ and the initial condition $P_1(0) = 1$.

In fig. 1 the formation of a self trapped state for $\chi \geq 4V$ is shown. The occupation difference oscillates between $x_3 = -1$ and 1 with mean value zero for $\chi < 4V$. For $\chi > 4$ it remains in the vicinity of $x_3 = 1$, i. e. the exciton is preferentially located at the site 1 of the dimer. For exactly $\chi = 4V$ the initial condition $P_1(0) = 1$ is located on the separatrix between trapped and detrapped states and approaches a very unstable hyperbolic fixed point [45] (due to the numerical inaccuracies the curve actually shown corresponds to $\chi$ a little bit below 4.0).

For both, $\gamma$ and $\chi$ different from zero we show in the parts (a) of the figs. 2 - 6 the time dependence of the total occupation probability $n(t)$ and in the parts (b) the occupation difference $x_3(t)$ rescaled with the total occupation probability.

In fig. 2 the influence of the small trapping rate $\gamma = 0.5$ is displayed. The total occupation probability decays slowly. Strong nonlinearity ($\chi \geq 5$) leads to an effective "switching off" of the trapping. In this case the probability remains almost constant until $t > 10/V$.

In figs. 3 and 4 the medium trapping rates $\gamma = 1., 2.$, respectively, have been chosen. For the linear dimer $\chi = 0$ the life time of the exciton is smaller for $\gamma = 2$ than for large trapping rates $\gamma = 5, 10$ (figs. 5, 6). This is due to the "fear of death effect" which becomes pronounced for $\gamma > 2$. It leads to a very large difference in the occupation probabilities of the sites 1 and 2. In fig. 7 the influence of different trapping rates (indicated at the right) on the occupation probability at the trap site $P_2(t)$ are presented for the linear dimer. $P_2(t)$ becomes smaller for larger trapping rate and the decay with time of the whole occupation probability is consequently slower.

A concise inspection of the probability difference $P_1(t) - P_2(t)$, scaled to the whole occupation probability (figs. 2 - 6 (b)) reveals, that for growing trapping strength the transition to the self trapped state takes place for $\chi < 4V$ already (but compare Fig. 2(b) in [32]). It is formed at the beginning of the time development and after some time it disappears. The self trapped state is not stable because the whole probability decays completely $P(t) \to 0$ for $t \to \infty$. Self trapping leads to a considerable enhancement of the life time of the exciton, e. g. for medium trapping rates (figs. 3, 4) and strong nonlinearity ($\chi = 6$) it is approximately four times longer than in the linear case $\chi = 0$.

A very strong trapping rate destroys the influence of the nonlinearity almost completely (figs. 5, 6). The transfer probabilities for the exciton are reduced so much, that the occupation probability $P_2(t)$ is very low and the decay of the whole occupation probability $P(t)$ is almost independent of the nonlinearity parameter $\chi$. However, the formation of a self trapped state is still reflected in the normalized occupation probability for short times.

## 4 The trimer with a sink

### 4.1 Previously obtained results

In the past some attention has been given to the solution of the nonlinear Schrödinger equation for the trimer. Preferentially symmetric trimers with symmetric initial conditions have been studied.

Eilbeck with his collaborators investigated the stationary solutions of the symmetric cyclic trimer. Their results are presented in Fig. 2 in [29]. They derived a bifurcation point and discussed the stability of the obtained solutions.

Kenkre with his collaborators were able to obtain an equation for the probability difference $x_3 = p(t) = P_1(t) - 2P_2(t)$ for a symmetric trimer with symmetric initial conditions [39,40].

## 4.2 Equations of motion

We consider a trimer with $V_{12} = V_{21} = -V$ and $V_{23} = V_{32} = -V' < 0$. A sink is introduced at site 3. Similarly to the case of the dimer one can pass to generalized Bloch variables $x_{1-8}$ containing also the variables $x_{1-3}$ which were previously introduced for the dimer. The procedure of introducing the generalized variables $x_{1-8}$ is explained in the appendix. One obtains

$$
\begin{aligned}
x_1 &:= \rho_{12} + \rho_{21} \\
x_2 &:= i(\rho_{12} - \rho_{21}) \\
x_3 &:= \rho_{11} - \rho_{22} \\
x_4 &:= \rho_{31} + \rho_{13} \\
x_5 &:= i(\rho_{13} - \rho_{31}) \\
x_6 &:= \rho_{23} + \rho_{32} \\
x_7 &:= i(\rho_{23} - \rho_{32}) \\
x_8 &:= \rho_{11} - \rho_{33} \\
x_9 &:= \rho_{22} - \rho_{33} \\
n &:= \rho_{11} + \rho_{22} \\
N &:= \rho_{11} + \rho_{22} + \rho_{33}
\end{aligned} \tag{18}
$$

Besides the linearly independent variables we have also introduced $x_9 = x_8 - x_3$ as well as the occupation sum of the sites 1 and 2 $n$ and the total trimer occupation $N$. The variable $n$ will be useful for considering the sink influence at site 3. We obtain the generalized system of Bloch equations

$$
\begin{aligned}
\dot{x}_1 &= +(\epsilon_2 - \epsilon_1)x_2 - V'x_5 \\
\dot{x}_2 &= +(\epsilon_1 - \epsilon_2)x_1 + V'x_4 + 2Vx_3 \\
\dot{x}_3 &= -2Vx_2 + V'x_7 \\
\dot{x}_4 &= +(\epsilon_3 - \epsilon_1)x_5 + Vx_7 - V'x_2 - (\gamma/2)x_4 \\
\dot{x}_5 &= +(\epsilon_1 - \epsilon_3)x_4 - Vx_6 + V'x_1 - (\gamma/2)x_5 \\
\dot{x}_6 &= +(\epsilon_3 - \epsilon_2)x_7 + Vx_5 - (\gamma/2)x_6 \\
\dot{x}_7 &= +(\epsilon_2 - \epsilon_3)x_6 - Vx_4 + 2V'x_9 - (\gamma/2)x_7 \\
\dot{x}_8 &= -Vx_2 + V'x_7 - (\gamma/3)(N - x_8 - x_9) \\
\dot{x}_9 &= +Vx_2 + 2V'x_7 - (\gamma/3)(N - x_8 - x_9) \\
\dot{n} &= -V'x_7 \\
\dot{N} &= -(\gamma/3)(N - x_8 - x_9)
\end{aligned} \tag{19}
$$

## 4.3 Analytical results for strong trapping

According to (7), (8) strong trapping leads to

$$\rho_{13} = \frac{1}{\varepsilon_1 - \varepsilon_3 - i\gamma}(V\rho_{23} - V'\rho_{12}) \tag{20}$$

$$\rho_{23} = \frac{1}{\epsilon_2 - \epsilon_3 - i(\gamma/2)}(V'\rho_{33} - V'\rho_{22} + V\rho_{13}) \tag{21}$$

and

$$\rho_{33} = \frac{iV'}{\gamma}(\rho_{23} - \rho_{32}). \tag{22}$$

This means that all density matrix elements with one index 3 are of the order $O(V/\gamma)$ and therefore small, $\rho_{33}$ is even of order $O(V^2/\gamma^2)$. Keeping only terms of leading order one can express $x_{4-7}$ through $x_{1-3}$ and obtains a closed system of equations for the latter:

$$\begin{aligned}
\dot{x}_1 &= -(\tilde{\gamma}/2)\, x_1 + (\epsilon_2 - \epsilon_1)x_2 \\
\dot{x}_2 &= -(\tilde{\gamma}/2)\, x_2 - (\epsilon_2 - \epsilon_1)x_1 + 2Vx_3 \\
\dot{x}_3 &= -(\tilde{\gamma}/2)\,(x_3 - n) - 2Vx_2 \\
\dot{n} &= -(\tilde{\gamma}/2)\,(n - x_3)
\end{aligned} \tag{23}$$

Comparing this to (12) we notice that the equations of motion for the trimer reduce for very strong trapping to the dimer equations with a newly defined sink rate

$$\tilde{\gamma} := \frac{(2V')^2}{\gamma}. \tag{24}$$

This eq. (24) once again demonstrates the fear of death effect: The effective sink rate $\tilde{\gamma}$ becomes small compared to $V'$ for $\gamma \gg V'$, i. e. the sink site in the trimer is effectively decoupled.

## 4.4 Numerical results

We have calculated numerically the time dependence of the site occupation probabilities in the presence of trapping and of a nonlinearity. With respect to the nonlinearity we consider two different cases:

Configuration (I): The nonlinearity $\chi$ is the same for all three sites

Configuration (II): The sink site is linear, i. e. $\chi_1 = \chi_2 = \chi$, $\chi_3 = 0$

Our data were obtained through a numerical integration of the nonlinear Schrödinger equation (1). We have chosen for all figures $V = 1$ and the initial condition $P_1(0) = 1$. The sink influences site 3. In the analysis we will concentrate on self trapping, the fear of death effect and regular vs. chaotic time dependence.

For reference we show first the linear trimer without trap fig. 8a. In this case the exciton coherently oscillates between the ends of the trimer. The sites 1 and 3 have equal average occupation probabilities which are due to the special initial condition $P_1(0) = 1$ larger than the occupation of site 2. The situation is similiar in the case of a small trapping rate $\gamma = 0.2$ (fig. 8b) but the total occupation probability in this case slowly decays. Increasing the trap strength up to an intermediate value $\gamma = 2$ (fig. 8c) makes the decay faster until for $\gamma$ larger than $\sim 2$ the fear of death effect becomes pronounced and the life time of the exciton grows again but now with the occupation probability at site 3 much smaller than for small trapping rates. We show for example the situation at $\gamma = 10$ (fig. 8d) when the exciton oscillates between the sites 1 and 2 and the trap site is almost decoupled.

A moderate nonlinearity (up to $\chi \sim 2$) does not change the appearance of the graphs very much (fig. 9). Again the fear of death effect controls the behaviour for larger trapping rate and in this case there is not much difference between the two configurations (I) and (II, not displayed) because the occupation of site 3 is reduced so much. In contrast, for weak trapping the regular oscillations of type (I) change into a quite irregular behaviour as one removes the nonlinearity at site 3 (figs. 9a and b). This effect is also present without the trap (not displayed) and is obviously due to the destruction of the symmetry between the sites 1 and 3.

As one increases the nonlinearity further, a significant change of the behaviour towards a self trapped solution occurs. We discuss this first for the symmetric trimer of configuration (I) without trap. The regular behaviour up to $\chi = 2$ (fig. 10a) is replaced by a rather chaotic one for $\chi = 3$ (fig. 10b). Still the average occupation probabilities are approximately the same for all three sites. There is, however, a large intervall where the occupation of site 3 is very small. For $\chi = 4$ (10c) the transition to a self trapped state on the dimer has occured and the occupation of site 3 remains small all the time. The appearance of the time dependence is much more regular than close to the transition point in fig. 10b. A further increase of the nonlinearity parameter causes the exciton to be completely localized at site 1 (fig. 10d) for $\chi \geq 5$, after that no other qualitative changes occur, just frequency and amplitude of the small oscillations in the occupation probabilities slightly depend on $\chi$ (not displayed). The consecutive transitions can be explained by the bifurcations which the stationary solutions undergo as the nonlinearity changes. They have been discussed in [29] in detail for the nonlinear Schrödinger equation without sinks.

The following figure 11 shows that the bifurcations are still present for a small trap strength $\gamma = 0.2$ (cmp. also figs. 8b; 9a, b). In the symmetric conf. (I) the transition to self trapping on the dimer occurs for all sink rates later than in conf. (II). We illustrate this with the figs. 11a, b. The effect can be understood having in mind that equal site occupation probabilities for the asymmetric configuration (II) cause a shift in the effective site energies $\epsilon_i$ on the dimer which in turn obstructs the transition from the dimer to site 3. The transition to self trapping on site 1 is not affected by the two different configurations (not displayed).

Even for very large $\gamma$ there is a clear transition from self trapping on the dimer to a solution completely localized at site 1 (figs. 8d; 9d and 12). In this case, however, the occupation probability on site 3 is always small. This is now not due to nonlinearity induced self trapping but to the fear of death effect. In general, both effects work together in slowing down the decay of the excitation. The life time is e. g. for $\gamma = 10$ and $\chi = 5$ doubled in comparison to $\chi = 3$.

The formation of self trapped states due to nonlinearity is not stable for $\gamma > 0$: as the the total occupation probability becomes small with increasing time the effect disappears and the site occupation probabilities level out, as it can be seen e. g. from fig. 11b ($t > 35$) and fig. 12b ($t > 30$). The nonlinearity effects must decrease for large time since the occupation probabilities entering the effective site energies (4) then become negligibly small.

In contrast, the lowering of the occupation $\rho_{33}$ due to strong trapping does not depend on time. The sink site is almost decoupled and the behaviour of $P_1$ and $P_2$ can be interpreted as a perturbed dimer (figs. 9d and 12a,b). We note the interesting effect, that the time dependence looks quite regular as well for $\chi \leq 3$ (fig. 9d) as for $\chi \geq 5$ (fig. 12b) but much more chaotic for $\chi = 4$. We relate this to the fact, that for $\chi = 4$ the initial condition $P_1(0) = 1$ is just on the homoclinic orbit (cmp. fig. 1) which encloses the self trapped states for the system without trap. Time evolution will therefore take the system in the perturbed dimer very close to the hyperbolic fixed point where it is very unstable and as a result one observes chaos (see [45] - [49]).

In the last fig. 13 we present a numerical check for eq. (24). We compare the time evolution for a trimer with the large sink rate $\gamma = 10$ (full lines) to a dimer with $\gamma = 0.4$ (dots) according to this equation. We find good agreement in particular for $\chi = 0$ (a). For $\chi = 3$ the result is still reasonable, but not as good as in (a) since (24) was derived also under the assumption $\gamma \gg \chi$.

## 5 Conclusions

In the present paper we studied the combined effect of trapping due to a sink and transport nonlinearity due to excitonic - vibronic interactions on the transfer dynamics and the life time of an exciton. The specific systems we have analyzed were the dimer and the trimer with a chain topology. A sink was introduced at one end of the configuration using the formalism given in [28]. The initial condition for the numerical integration of the NLSE was always a state localized at the end opposite to the sink site.

The main effect of the nonlinearity is self trapping. In the dimer with sink a tendency to form a self trapped state could be observed for $\chi < 4V$, i.e. for smaller nonlinearity than in the isolated dimer. In the trimer self trapping on the sinkless subdimer for an intermediate nonlinearity was found besides states which are localized on one site only for strong nonlinearity. The transition to a self trapped state was accompanied by relatively irregular time dependencies. Though self trapping was enhanced by the sink for short times the decay of the total occupation probability destroyed the localized states for long times. Self trapping was shown to have a crucial influence on the life time of the excitation which grows for increasing nonlinearity.

For large sink rates the fear of death effect controls the decay of the excitaton. In this case the occupation of the sink site is reduced so much that the life time even increases with growing sink rate. The sink site is then almost decoupled from the rest as it was shown using an asymptotic approximation.

For small trapping rates there was a clear change in the behaviour when the nonlinearity of the sink site was was dropped while this effect was supressed by the effective decoupling of this site for strong trapping.

# Acknowledgements

Support by the Deutsche Forschungsgemeinschaft (DFG) is gratefully acknowledged. One of us (I. B.) wishes to thank the Deutscher Akademischer Austauschdienst (DAAD) for support and the Humboldt University Berlin for the kind hospitality.

# Appendix: Generalized Bloch representation for the trimer

The Bloch variables for the dimer are derived from the state vector $\underline{c} = (c_1, c_2)^T$ by employing the Pauli matrices $\sigma_i$

$$\sigma_1 = \begin{pmatrix} 0 & 1 \\ 1 & 0 \end{pmatrix}, \quad \sigma_2 = \begin{pmatrix} 0 & -i \\ i & 0 \end{pmatrix}, \quad \sigma_3 = \begin{pmatrix} 1 & 0 \\ 0 & -1 \end{pmatrix} \tag{25}$$

which are the generators of the Lie algebra SU(2). In this case one defines

$$x_i := \underline{c}^+ \sigma_i \underline{c} \qquad (i = 1, 2, 3). \tag{26}$$

Analogously, one derives generalized Bloch variables for the trimer from the state vector $\underline{c} = (c_1, c_2, c_3)^T$ using the Gell - Mann matrices $\lambda_i$, which are generators of the Lie algebra SU(3), in the form

$$\lambda_1 = \begin{pmatrix} 0 & 1 & 0 \\ 1 & 0 & 0 \\ 0 & 0 & 0 \end{pmatrix}, \quad \lambda_2 = \begin{pmatrix} 0 & -i & 0 \\ i & 0 & 0 \\ 0 & 0 & 0 \end{pmatrix}, \quad \lambda_3 = \begin{pmatrix} 1 & 0 & 0 \\ 0 & -1 & 0 \\ 0 & 0 & 0 \end{pmatrix}$$

$$\lambda_4 = \begin{pmatrix} 0 & 0 & 1 \\ 0 & 0 & 0 \\ 1 & 0 & 0 \end{pmatrix}, \quad \lambda_5 = \begin{pmatrix} 0 & 0 & -i \\ 0 & 0 & 0 \\ i & 0 & 0 \end{pmatrix}, \quad \lambda_6 = \begin{pmatrix} 0 & 0 & 0 \\ 0 & 0 & 1 \\ 0 & 1 & 0 \end{pmatrix} \quad (27)$$

$$\lambda_7 = \begin{pmatrix} 0 & 0 & 0 \\ 0 & 0 & -i \\ 0 & i & 0 \end{pmatrix}, \quad \lambda_8 = \begin{pmatrix} 1 & 0 & 0 \\ 0 & 0 & 0 \\ 0 & 0 & -1 \end{pmatrix}, \quad \lambda_9 = \begin{pmatrix} 0 & 0 & 0 \\ 0 & 1 & 0 \\ 0 & 0 & -1 \end{pmatrix}$$

They generate the variables

$$x_i := \underline{c}^+ \lambda_i \underline{c} \qquad (i = 1, \ldots, 9). \quad (28)$$

These variables $x_i$ are by definition real. We have chosen $\lambda_8$ in a modified form convenient for our calculations and added $\lambda_9$ which is linearly dependent ($\lambda_9 = \lambda_8 - \lambda_3$). Our variables $x_8$ and $x_9$ are therfore generalizations of the variable $x_3$ and have the form of differences between occupation probabilities.

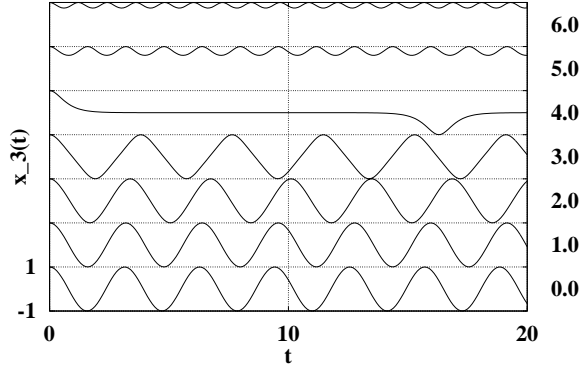

**Figure 1:** *Occupation probability difference $x_3(t)$ for the trapless dimer ($\gamma = 0$) and various nonlinear coupling strengths $\chi$ (given to the right of the plots). The transition to a self trapped state at site 1 ($x_3 \sim 1$) is obvious.*

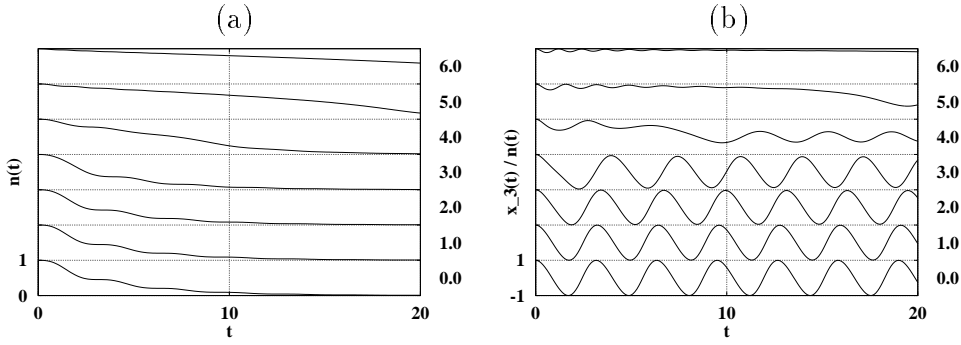

**Figure 2:** *Total occupation probability $n(t)$ (a) and normalized occupation probability difference $x_3(t)/n(t)$ (b) for a dimer with weak trapping $\gamma = 0.5$ and various nonlinear coupling strengths $\chi$ (given to the right of the plots). Again the formation of a self trapped state for $\chi \geq 4$ can be observed. It is however not stable and decays after some time ($t \sim 10$ for $\chi = 4$)*

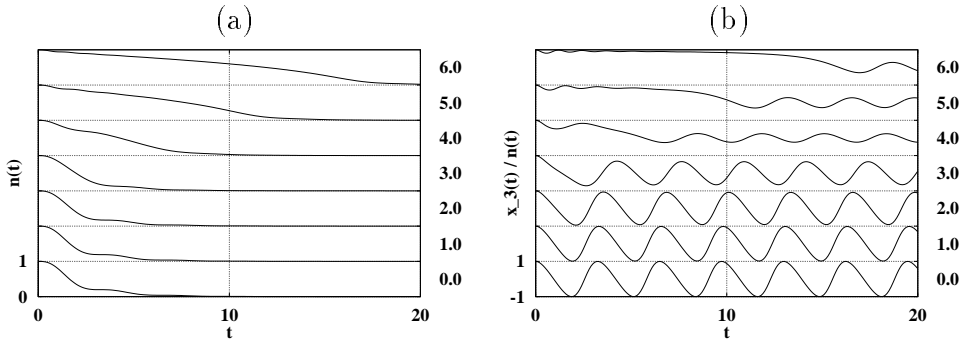

**Figure 3:** *Total occupation probability (a) and normalized occupation probability difference (b) for a dimer with intermediate trapping $\gamma = 1.0$. The decay of the total occupation is faster than for $\gamma = 0.5$ and also the self trapped state disappears earlier. The influence of self trapping on the life time of the exciton is evident.*

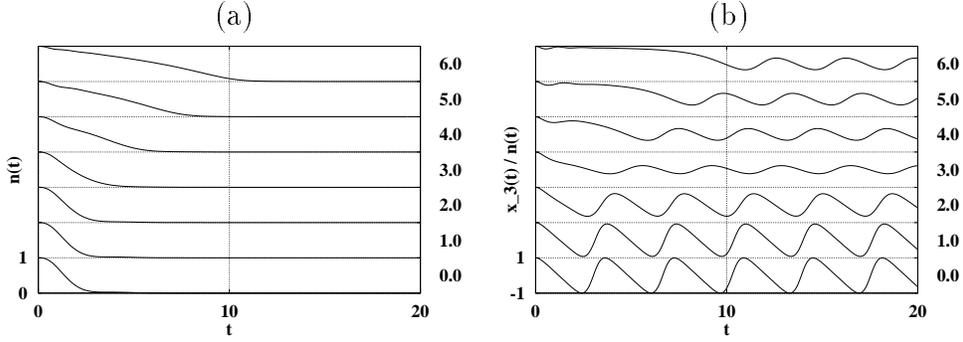

**Figure 4:** *Total occupation probability (a) and normalized occupation probability difference (b) for a dimer with intermediate trapping $\gamma = 2.0$. The tendency to form a self trapped state can be recognized even for $\chi = 3$.*

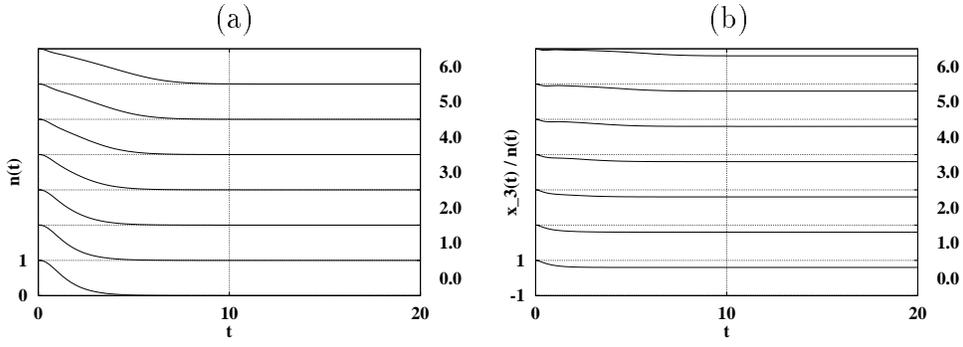

**Figure 5:** *Total occupation probability (a) and normalized occupation probability difference (b) for a dimer with strong trapping $\gamma = 5.0$. The life time of the exciton is similiar to the case $\gamma = 2$ but clearly shorter than for $\gamma = 10$. The oscillations of the occupation difference have disapeared and the occupation at site 1 is always much larger than at the sink site 2 (fear of death effect).*

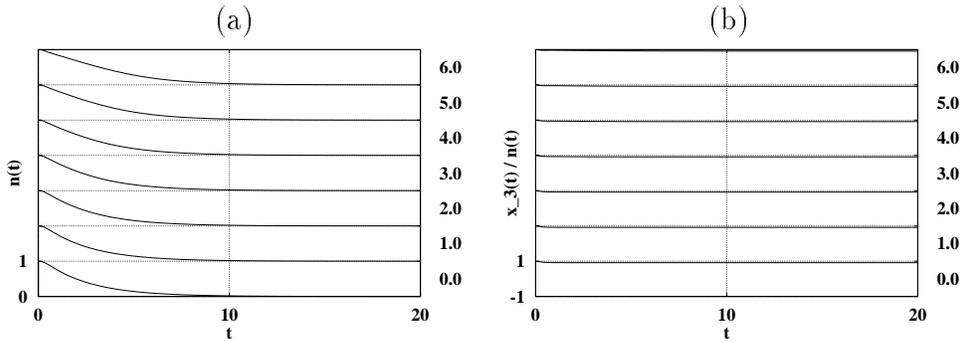

**Figure 6:** *Total occupation probability (a) and normalized occupation probability difference (b) for a dimer with very strong trapping $\gamma = 10$. The exciton is almost completely localized at site 1. The influence of the nonlinear coupling is very small.*

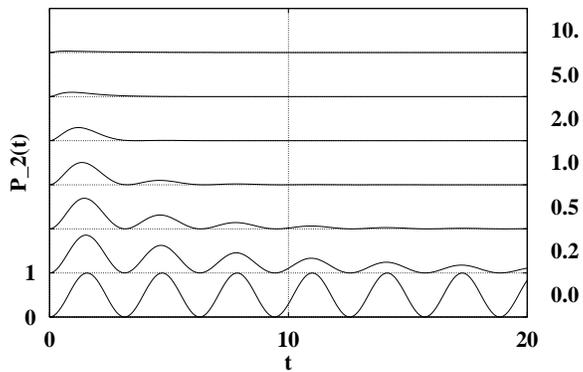

**Figure 7:***Occupation probability $P_2(t)$ of the site with the trap for no nonlinearity ($\chi = 0$) and various trapping strengths $\gamma$ (given to the right of the plots). The fear of death effect leads to the very small occupation of the sink site for $\gamma \geq 5$.*

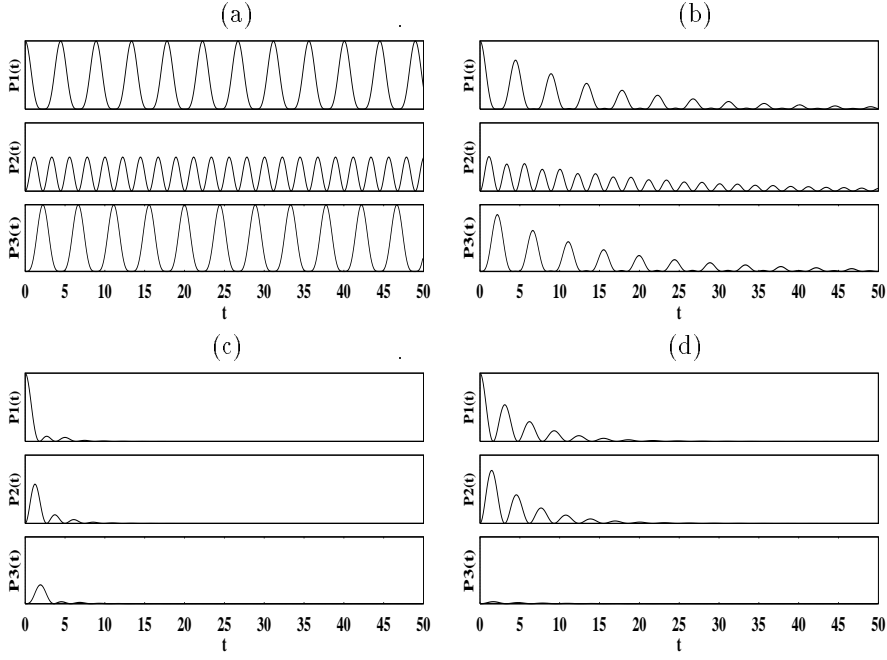

**Figure 8:** *Decay of an exciton on a linear trimer ($\chi = 0$) for different sink rates: (a) $\gamma = 0$, (b) $\gamma = 0.2$, (c) $\gamma = 2$, (d) $\gamma = 10$. The fear of death effect causes the life time of the exciton to be shorter for $\gamma = 2$ than for $\gamma = 10$.*

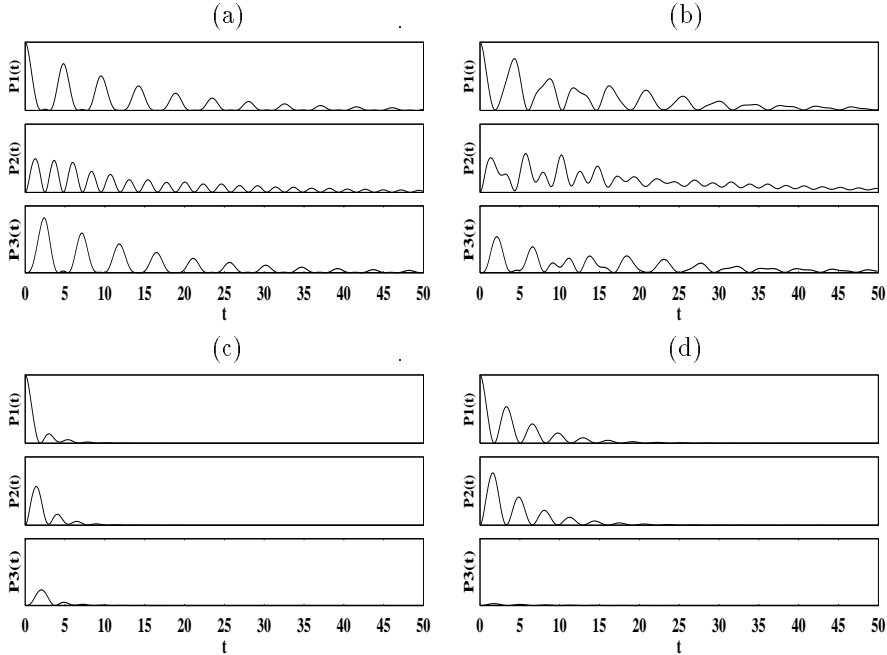

**Figure 9:** *Decay of an exciton on a trimer with small nonlinear coupling $\chi = 2.0$ for different sink rates and configurations: (a) $\gamma = 0.2$, conf. (I) $\chi = \chi_i (i = 1, 2, 3)$; (b) $\gamma = 0.2$, conf. (II) $\chi_1 = \chi_2 = \chi$, $\chi_3 = 0$; (c) $\gamma = 2$, (d) $\gamma = 10$, both conf. (I). Conf. (II) shows irregular behaviour for small sink rate due to the destruction of symmetry (b). For larger sink rate conf. (II) is not displayed because it behaves very similiar to conf. (I). Comparison of (c) and (d) again shows the fear of death effect.*

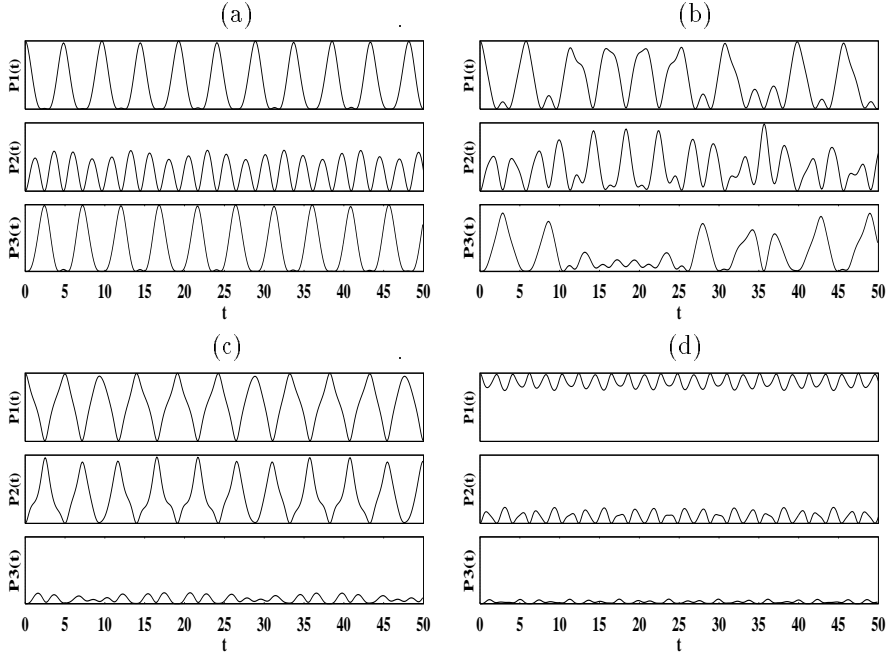

**Figure 10:** *The formation of a self trapped state for the trapless symmetric trimer (conf. (I), $\gamma = 0$): At $\chi = 2$ (a) the behaviour is very similiar to the linear trimer of fig. 8a. The time dependence becomes irregular for $\chi = 3$ (b). For $\chi = 4$ (c) the solution is self trapped on a dimer and for $\chi = 5$ (d) on site 1.*

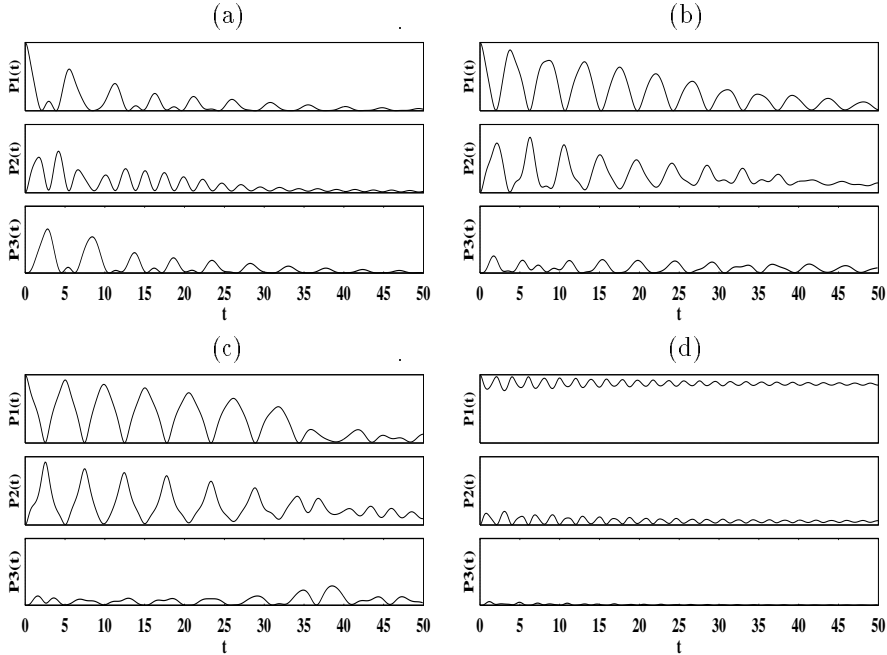

**Figure 11:** *Formation of a self trapped state for a trimer with the small sink rate $\gamma = 0.2$: (a) $\chi = 3$, conf. (I), (b) $\chi = 3$, conf. (II), (c) $\chi = 4$, and (d) $\chi = 5$, both conf. (I). For conf. (II) self trapping on the dimer occurs for smaller nonlinearities. The self trapped state on the dimer in (c) dissappears for $t \geq 35$.*

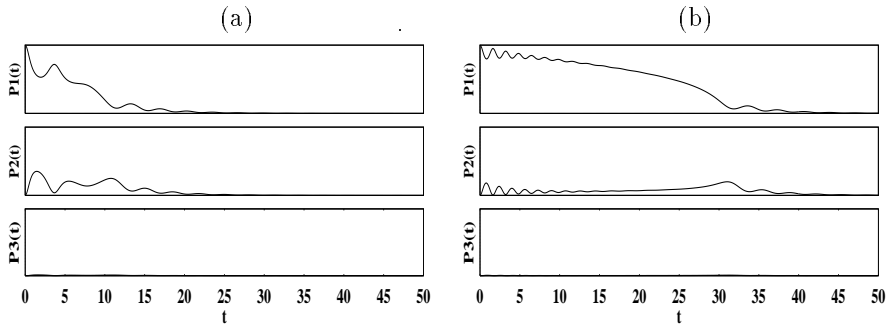

**Figure 12:** *Transition to a self trapped solution on site 1 for strong trapping $\gamma = 10$. The nonlinear coupling is (a) $\chi = 4$ and (b) $\chi = 5$. The self trapped state is unstable for $t \geq 30$. The occupation of site 3 is always very small (fear of death effect). The time dependence for $\chi = 4$ is very irregular.*

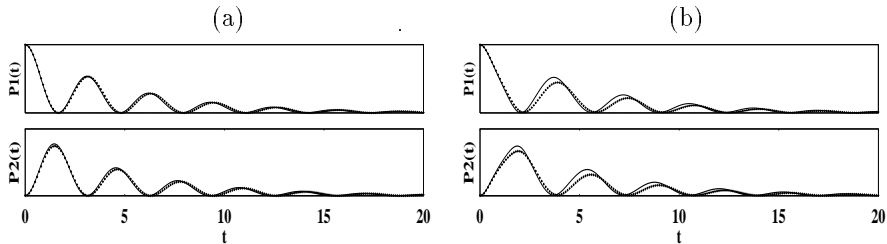

**Figure 13:** *Reduction of the trimer with strong trapping to an effective dimer with weak trapping. The full line shows the occupation probabilities of the sites 1 and 2 for a trimer with $\gamma = 10$ and (a) $\chi = 0$, (b) $\chi = 3$. The dots show according to eq. (24) the corresponding effective dimer with the sink rate $\gamma = 0.4$*